# Numerical simulation of magnetization process in antiferromagnetic-ferromagnetic bilayer with compensated interface

N. A. Usov
*Troitsk Institute for Innovation and Fusion Research, 142190, Troitsk, Moscow region, Russia*

S. A. Gudoshnikov
*Institute of Terrestrial Magnetism, Ionosphere and Radio Wave Propagation RAS (IZMIRAN),*
*142190 Troitsk, Moscow region, Russia*

The properties of antiferromagnetic (AFM)-ferromagnetic (FM) bilayer have been studied using self-consistent mean-field approximation for Heisenberg Hamiltonian. The perpendicular exchange coupling has been revealed in a bilayer with a compensated interface. For a uniform AFM film a symmetrical hysteresis loop has been calculated, because the transverse instability develops within the AFM film at certain critical value of external magnetic field. On the other hand, shifted hysteresis loop with a finite exchange bias field has been obtained for a non-uniform AFM film consisting of various AFM domains with perpendicular directions of the easy anisotropy axes.

PACS numbers: 75.30.Et, 75.30.Pd, 75.60.Ej, 75.70.Cn

## I. INTRODUCTION

The exchange coupling at the interface between ferromagnetic (FM) and antiferromagnetic (AFM) thin films can shift the hysteresis loop of the system along the magnetic field axis.[1-4] This phenomenon, the so-called exchange bias, has been extensively studied in recent years both experimentally and theoretically.[3-11] In particular, the numerical simulations of Koon[7], and Schulthess and Butler[8,9] showed the possibility of perpendicular coupling at a compensated AFM surface. The perpendicular coupling has been discovered in experiments on $FeF_2$/Fe and $MnF_2$/Fe bilayers[12-14] with (110) and (101) compensated interfaces, as well as in experiment on $Fe_3O_4$/CoO bilayer with compensated (001) CoO interface.[15-17] Nevertheless, the nature of the perpendicular coupling has not been fully understood so far.

It has been shown recently[18] that certain periodic modulation of FM and AFM spin orientations near a compensated interface of FM - AFM bilayer is energetically favorable. The amplitude of the modulation decreases exponentially into the AFM and FM volumes. Therefore, from a macroscopic point of view, the AFM - FM exchange interaction can be considered as a surface magnetic anisotropy, the interaction energy being proportional to a square of a scalar product of the unit FM and AFM vectors at the interface. Using an explicit expression for the interfacial interaction energy an analytical theory has been developed[19] for a uniform AFM - FM bilayer with a compensated interface. It has been shown that even for a high value of surface interaction strength a transverse instability originates within the uniform AFM film, because the deviation of the unit AFM vector out of the interface plane becomes energetically favorable in a large enough external magnetic field.

It should be pointed out, however, that in the real experiment on $FeF_2$/Fe and $MnF_2$/Fe bilayers[12-14], as well as for $Fe_3O_4$/CoO bilayer[15-17], the AFM film is certainly non-uniform. Rather, it consists of tiny domains with various directions of the easy anisotropy axes at the interface. To investigate this particular situation in the present paper a new numerical simulation scheme has been developed based on a self-consistent mean-field approximation for Heisenberg Hamiltonian.[20] The numerical simulations confirm the existence of a perpendicular exchange coupling at the compensated interface of the FM - AFM bilayer. It has been found also that for a uniform AFM film a transverse instability develops within the AFM film at a critical value of external magnetic field. This instability breaks the initial exchange coupling between FM and AFM layers resulting in a symmetrical hysteresis loop. On the other hand, shifted hysteresis loop with a finite exchange bias field has been obtained for a non-uniform AFM film consisting of various AFM domains with perpendicular directions of the easy anisotropy axes. The lateral domain sizes have to be comparable with the FM exchange length. In this case the conditions for the development of the transverse instability near the interface cannot be satisfied simultaneously for various domains. This leads to a stabilization of the AFM spin distribution even in a very high magnetic field.

## II. NUMERICAL SIMULATION MODEL

To study the magnetic properties of AFM-FM bilayer we use Heisenberg Hamiltonian

$$H = -g\mu_B \mathbf{h}_0 \sum_i \mathbf{S}_i - \frac{1}{2}\sum_{i \neq j} J_{ij} \mathbf{S}_i \mathbf{S}_j + \frac{1}{2}(g\mu_B)^2 \sum_{i \neq j}\left(\frac{\mathbf{S}_i \mathbf{S}_j}{R_{ij}^3} - \frac{3(\mathbf{R}_{ij}\mathbf{S}_i)(\mathbf{R}_{ij}\mathbf{S}_j)}{R_{ij}^5}\right) - \sum_i D_x (S_i^x)^2 ,$$

(1)

where $g$ is the Lande factor, $\mu_B$ is the Bohr magneton, $\mathbf{h}_0$ is the external magnetic field, $\mathbf{S}_i$ is the spin operator in the $i$-th lattice site. The first term in Eq. (1) is the Zeeman energy, the second term describes the exchange interactions between the nearest neighbors, $J_{ij}$ being the corresponding exchange integrals. The third term gives the energy of mutual dipole-dipole interactions, where $\mathbf{R}_{ij}$ is the vector connecting $i$ and $j$ lattice sites. The last term represents the one-ion anisotropy contribution. For uniform AFM film the easy anisotropy axis is assumed to be parallel to the $x$-axis, $D_x$ being the corresponding one-ion anisotropy constant.

In the Hartree – Fock approximation the eigen-function of the Hamiltonian (1) is represented by a product of one-particle functions in different lattice sites, $\Psi = \psi_1 \psi_2 .. \psi_N$. These functions are normalized by unity, $<\psi_i|\psi_i> = 1$. Using variational principle one verifies that the functions $\psi_i$ satisfy the following set of non-linear equations ($i = 1, 2... N$)

$$\left( -g\mu_B(\mathbf{h}_0 + \mathbf{h}_{mi})\mathbf{S}_i - \sum_{j \neq i} J_{ij}\langle \mathbf{S}_j \rangle \mathbf{S}_i - D_x(S_i^x)^2 \right)\psi_i = \varepsilon_i \psi_i, \quad (2)$$

where $<\mathbf{S}_j> = <\psi_i|\mathbf{S}_j|\psi_i>$, and the sum in the second term is over the nearest neighbors of the site $i$. The demagnetizing field at the site $i$ is given by

$$\mathbf{h}_{mi} = -g\mu_B \sum_{j \neq i} \left( \frac{\langle \mathbf{S}_j \rangle}{R_{ij}^3} - \frac{3\mathbf{R}_{ij}(\mathbf{R}_{ij}\langle \mathbf{S}_j \rangle)}{R_{ij}^5} \right).$$

Note, the anisotropy constant is usually small with respect to the characteristic value of the exchange integrals, $D/J \ll 1$, so that the anisotropy energy can be considered as a perturbation. Then, in the lowest approximation of the perturbation theory one obtains from Eq. (2) the Hamiltonian ($\alpha = x, y, z$)

$$H_i^{(0)} = -\sum_\alpha A_{i\alpha} S_i^\alpha;$$
$$A_{i\alpha} = g\mu_B(h_0^\alpha + h_{mi}^\alpha) + \sum_{j \neq i} J_{ij}\langle S_j^\alpha \rangle. \quad (3)$$

Here $A_{i\alpha}$ are the real coefficients having the meaning of the effective field components in the $i$-th lattice site (we will omit index $i$ further). Using the commutation relations for the components of the spin operators, i.e. $[S_x, S_y] = iS_z$, etc.[21], it is easy to construct for the Hamiltonian (3) the lowering and raising operators. Then, the eigen-functions of (3) can be numerated by means of quantum number $n = -S, -S +1, .. S$, $S$ being the spin length. The corresponding eigen-values are given by $E_n = n\Delta$, $\Delta = (A_x^2 + A_y^2 + A_z^2)^{1/2}$. Besides, the averaged values of the spin operators are easily calculated as $<n|S^\alpha|n> = A_\alpha E_n/\Delta^2$. With these relations the Hartree-Fock procedure becomes self-consistent, because the coefficients $A_\alpha$ are, in turn, the functions of the averaged values of the spin operators. Now the influence of the single-ion anisotropy can be taken into account considering the operator $D_x(S^x)^2$ as the perturbation. One can see therefore, that the numerical procedure described[20] for the simplest case $S = \frac{1}{2}$ can be easily generalized to the case of arbitrary $S$ value.

Having in hand the eigen-functions and eigen-values of the Hamiltonian (1) one can calculate expectation values of the spin operators in different lattice sites. This enables one to determine various physical properties, in particular, average magnetization of the spin system as function of external magnetic field.

To apply this approach to AFM-FM bilayer one can consider Hamiltonian (1) on a simple cubic lattice $i = (n, m, k)$. Let upper $N_f$ plates, $k = 1, 2, .. N_f$, contain FM spins with exchange integral $J_{ij} = J_f > 0$, whereas lower $N_a$ plates, $k = -N_a+1, -N_a+2, .., 0$ contain AFM spins, with $J_{ij} = J_a < 0$ respectively. The exchange integral $J_{int}$ of arbitrary sign describes the exchange interaction between the sites of the top AFM and the bottom FM plates. In the following calculations we adopt different anisotropy constants, $D_f \ll D_a$, for FM and AFM films, respectively.

### III. NATURE OF SURFACE INTERACTION

Based on the Hamiltonian (1) it has been proved analytically[18] that certain orientation modulation of the AFM and FM spin structure near a compensated surface of AFM film is energetically favorable. Neglecting the demagnetizing and anisotropy energy contributions as small corrections, the interaction energy per unit square of the AFM-FM interface has been obtained[18] as

$$E = -K_s \sin^2 \varphi;$$
$$K_s = \frac{J_f S_f^2 + J_a S_a^2}{17.6 J_f J_a a^2} J_{int}^2, \quad (4)$$

where $K_s$ has the meaning of the surface anisotropy constant, $\varphi$ is the angle between the unit FM and AFM vectors, $S_f$ and $S_a$ are the FM and AFM spin lengths, respectively, $a$ being the lattice period. The surface anisotropy constant was roughly estimated[18] to be $K_s \sim 1$ erg/cm$^2$, in reasonable agreement with the experimental values[3]. It follows from Eq. (4) that the interaction energy is minimal at $\varphi = \pi/2$, so that the perpendicular orientation of the AFM and FM spins is preferable at the AFM-FM interface with compensated AFM surface.

The result (4) has been obtained[18] in the quasi-classical approximation, $S_f, S_a \gg 1$. Besides, the demagnetizing and anisotropy energy contributions to the Hamiltonian (1) were omitted in comparison with the main exchange energy contribution. The approach described in Section 2 enables one to calculate the actual spin structure near the AFM-FM interface avoiding the limitations mentioned above. Fig. 1 shows the top view of the calculated spin distributions in the bottom FM plate, $k = 1$, and the top AFM plate, $k = 0$, respectively in zero magnetic field. Spin deviations are negligibly small for the FM plates with $k > 1$ and AFM plates with $k < 0$. It is easy to see that the spin structures shown in Fig 1 are consistent with the analytical expressions given by Eq. (4) in Ref.[18] Only a small portion of the area interface is presented in Fig. 1 because the sample contains totally 24×24×30 spins, with 24 plates of AFM spins and 6 upper plates of FM spins. The Heisenberg Hamiltonian parameters were assumed to be: spin values $S_f = S_a = 1$; exchange integrals $J_f = 100k_B$, $J_a = -60k_B$ and $J_{int} = 200k_B$, where $k_B$ is the Boltzmann constant; anisotropy constants $D_f = 0.05k_B$ and $D_a = 3k_B$, lattice period $a = 4 \cdot 10^{-8}$ cm. The exchange integral $J_{int}$ has been chosen large enough to make the spin deviations in Fig. 1 clearly visible. Both AFM and FM easy anisotropy axes are assumed to be parallel to the $x$-axis. Nevertheless, one can see in the Fig. 1 that for the ground state of the bilayer the average direction of the FM spins is parallel to the $y$-axis. Therefore, FM spins rotate perpendicular to the AFM spins due to the influence of the exchange interaction at the interface.

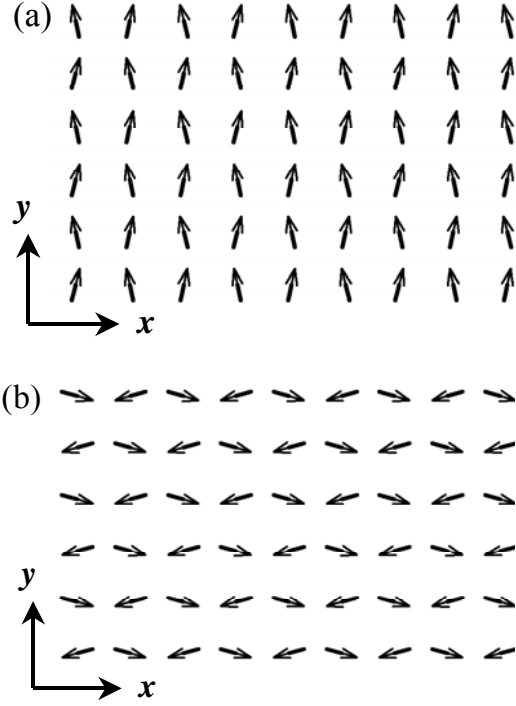

FIG. 1. Spin configurations in the bottom FM plate (a) and the top AFM plate (b) in the absence of external magnetic field.

## IV. TRANSVERSE INSTABILITY

Using the explicit expression (4) for the interfacial interaction energy density the analytical theory for AFM-FM bilayer with a compensated interface has been developed[19]. This investigation shows that the properties of the bilayer depend crucially on the surface interaction strength $\eta = K_s/\delta K_a$, where $K_a$ is the anisotropy constant of the AFM film, $\delta = (C_a/2K_a)^{1/2}$ has the meaning of the AFM domain wall width, $C_a$ being the AFM exchange constant. If surface interaction strength is small, $\eta < 2$, the exchange coupling at the interface breaks in a moderate magnetic field, because the rotation of the unit AFM vector at the interface is restricted by a critical angle, $\psi_{max} = \arcsin(\eta/2)$. As a result the bilayer exhibits usual hysteretic behavior in an external magnetic field. However, its coercive force increases with respect to that of free FM film due to the influence of the interface interaction. On the other hand, if $\eta \geq 2$ a formal 1D solution exists in arbitrary high external magnetic field. It describes purely reversible behavior of the bilayer similar to that presented by the Koon's model[7]. This is because the energy stored in AFM and FM domain walls returns back when external magnetic field decreases to zero. However, it has been found[19] that transverse instability happens in AFM film at certain critical magnetic field, when a deviation of the unit AFM vector out of the interface plane becomes energetically favorable. This instability is similar to that one discussed by Brown many years ago[22].

The conclusions obtained in[19] are consistent with the results of numerical simulation for a bilayer with a uniform AFM film. The curve a) in Fig. 2 shows the symmetrical hysteresis loop calculated for a sample consisting of 30 plates of AFM spins augmented by 6 plates of FM spins. Each plate contains 30 × 30 interacting spins. The material parameters are the same as in Fig, 1 with the difference that $J_a = -40k_B$ and $J_{int} = 80k_B$, so that the surface anisotropy constant $K_s = 2.27$ erg/cm$^2$. To account for a large value of the saturation magnetization of the FM film, the Lande factors are assumed to be $g_a = 2.0$ and $g_f = 4g_a$, for the AFM and FM spins, respectively. Due to comparatively large value of the AFM anisotropy constant, in the ground state the AFM spins are parallel to their easy anisotropy axis ($x$-axis). Then, to minimize the interaction energy (4) the FM spins rotate perpendicular to the $x$-axis. In principle, the interaction energy (4) is invariant with respect to the rotation of the FM spins within the $yz$ plane. However, due to the shape anisotropy, for a thin FM film the FM spins are restricted within the interface plane, being parallel to the $y$-axis. Therefore, the initial configuration in zero magnetic field is similar to that one shown in Fig. 1. Of course, there is another equivalent configuration when FM magnetization points along negative direction of the $y$-axis.

For the loop a) in Fig. 2 the external magnetic field is applied within the interface plane at the angle $\varphi_H = 4\pi/3$ with respect to the $x$-axis. Therefore, it is nearly opposite to the initial direction of the FM magnetization. The FM magnetization rotates gradually towards the magnetic field direction when applied magnetic field increases.

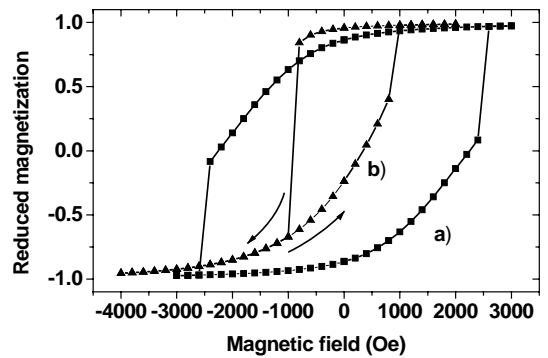

FIG. 2. Hysteresis loops for different AFM-FM bilayers; a) uniform AFM film; b) AFM film consists of random AFM domains with perpendicular easy anisotropy axes.

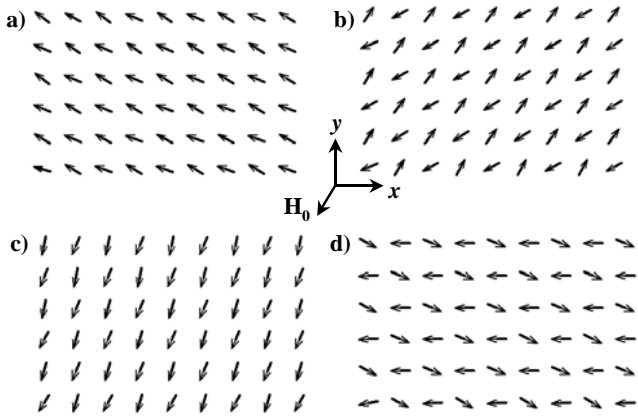

FIG. 3. Spin configurations in the bottom FM and the top AFM plates before and after the transverse instability. a), b) panels: $h_0 = 2400$ Oe; c), d) panels: $h_0 = 2600$ Oe.

The AFM spins follow this rotation trying to be nearly perpendicular to the FM spin direction at the interface. As a result, the domain walls originate within the FM and AFM films (see Fig. 1 in Ref.[19]). Panels a) and b) in Fig. 3 shows spin arrangements for the FM and AFM plates closest to the interface at $h_0 = 2400$ Oe, just before the magnetization reversal point at the hysteresis loop a) in Fig. 2. Up to this field the spin arrangements for all plates remain parallel to the interface plane. The characteristic modulation of the tilted spin structure is clearly visible in the panel b) of the Fig. 3. It exists also for the panel a) of this figure though its amplitude is small. Again the average directions of the FM and AFM spins at the interface are nearly perpendicular to each other. Thus, the AFM spins rotate within the interface plane at a large angle from the easy axis direction.

Note that the rotations of the AFM and FM spins are reversible up to $h_0 = 2400$ Oe, because in the absence of any imperfections the energy stored within the FM and AFM domain walls returns back if external magnetic field decreases to zero. The reversible rotation of the FM and AFM spins would lead to the exchange biasing if the AFM-FM spin system were stable during the evolution in external magnetic field up to high enough values of $h_0$. However, at $h_0 = 2600$ Oe the transverse instability[19] happens within the AFM film, because at this value of applied magnetic field the AFM domain wall loses its stability. As a result, the AFM spins deviate perpendicular to the interface plane and initial exchange coupling between the AFM and FM films breaks. The stable spin configurations for the FM and AFM plates closest to the interface at $h_0 = 2600$ Oe are shown in the panels c) and d) of Fig. 3, respectively. One can see that during transverse instability the FM spins rotate irreversibly to another stable direction close to negative direction of the y-axis, whereas the AFM spins return to their easy axis. Besides, all spins are parallel to the interface plane.

It can be proved[19] that the transverse instability is inevitable for AFM-FM bilayer with uniform AFM film having a uniaxial type of magnetic anisotropy. Due to the transverse instability the energy stored within the AFM and FM domain walls disappears. Therefore, for the AFM-FM bilayer with the uniform AFM film the exchange bias field is zero. The corresponding hysteresis loop is symmetrical with respect to the magnetic field axis similar to the loop a) shown in Fig. 2, though the coercive force of AFM-FM bilayer increases considerably with respect to that of free FM film.

## V. AFM DOMAIN MODEL

It follows from the above discussion that a finite exchange bias field can be obtained for AFM-FM bilayer only if the transverse instability at its interface is somehow avoided. As we have already mentioned, in the real experiment the AFM film consists of tiny twin domains. The directions of the easy anisotropy axes in different domains are perpendicular to each other at the interface.[13,14] Therefore, even if the condition for the onset of the transverse instability is fulfilled for the one type of domains, it cannot be satisfied for the other type of domains having perpendicular direction of the easy anisotropy axis. One can assume that this leads to a stabilization of the spin distribution in the AFM film near the interface.

To model the experimental situation let us consider a non-uniform AFM film consisting of various domains with perpendicular directions of the easy anisotropy axes. For the same sample discussed in the Section 4 we divide the AFM film consisting of 30×30×30 spins into 108 rectangular domains having 5×5×10 spins each. The easy anisotropy axis within a domain is parallel to either x or y-axis with the one-ion anisotropy constants $D_{ax} = D_{ay} = 3k_B$. The directions of the easy axes are distributed randomly among the domains. Fig. 4 shows typical distribution of the domains' easy axes at the AFM – FM interface. The dark and white areas represent the domains where the easy anisotropy axes are parallel to the *x*- and *y*-axis, respectively. In fact, the adjacent domains with the same easy axis direction join together, so that the random configuration in Fig. 4 consists of various domains of irregular shapes with different lateral sizes. The lateral size of the initial domains is given by $5a$ = 2 nm, so that the actual lateral domain size can be several times larger. Nevertheless, it is of the order of the exchange length of the FM film. Therefore, the exchange interaction within the FM film is sufficiently strong to keep the directions of the FM spins nearly parallel to each other even close to the interface. On the other hand, for the AFM spins there is an effective easy anisotropy axis. Its direction is tilted at 45° to the *x*-axis direction. This situation is demonstrated in Fig. 4, where the average directions of the unit AFM vectors in the domains are shown in zero magnetic field. The exchange interaction between the FM and AFM spins rotates now the FM magnetization perpendicular to the effective AFM easy axis. Therefore, in zero magnetic field the FM magnetization is tilted approximately at 135° to the *x*-axis direction, if the volume fractions of the *x*- and *y*-AFM domains at the interface are nearly equal. This spin distribution is in agreement with the experimental data.[14]

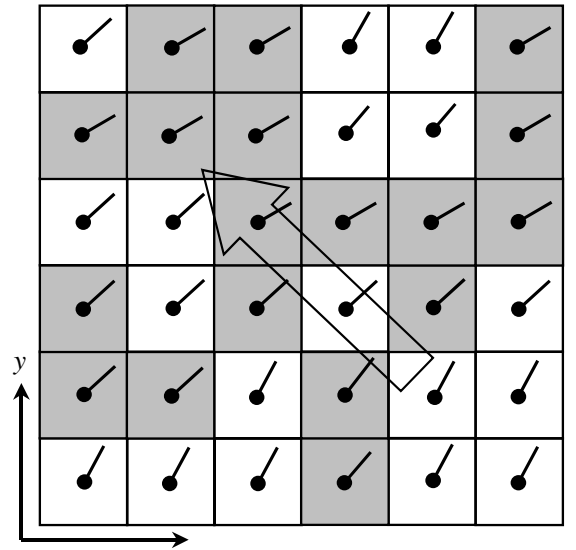

FIG. 4. Random AFM domains at the bilayer interface. Small arrows show averaged directions of the unit AFM vectors in various domains, large arrow represents the direction of FM magnetization. For the dark and white areas the AFM easy axis is parallel to x and y axes, respectively.

The calculated hysteresis loop for this sample is given by the curve b) in Fig. 2. The external magnetic field is applied within the interface plane at the angle $\varphi_H$ = 5π/3 with respect to the *x*-axis, to be nearly opposite to the initial direction of the FM magnetization. Importantly, the transverse instability *is absent* for the sample with non-uniform AFM film. At applied magnetic field $h_0$ = 1000 Oe the abrupt rotation of the FM magnetization starts, being accompanied by the rotation of the unit AFM vectors in the AFM domains. But this rotation is *partly* reversible, because the AFM – FM coupling at the interface does not break completely even in a very large applied magnetic field, up to $h_0$ = 4000 Oe. As a result, on the way back, along the lower part of the hysteresis loop b) in Fig. 2, the component of FM magnetization along the magnetic field direction vanishes not at $h_0$ = 1000 Oe, but at much lower field, $h_0$ = 400 Oe. Therefore, according to the usual definition of the exchange bias field[3], for the loop b) in Fig. 2 it is given by $h_{eb}$ = 300 Oe. Of course, the shape of the loop obtained is rather unusual, but this may be related with various factors, such as actual distribution of the domain sizes and anisotropy constants for different AFM domains. The AFM domain model demonstrates the important role of AFM domains in the exchange biasing. Actually, it is well recognized now[23,24] that it is the rearrangement of the AFM domains during the rotation of the FM magnetization that makes the experimental situation for $FeF_2$/Fe and $MnF_2$/Fe bilayers so complicated.

## VI. CONCLUSIONS

The numerical simulation based on self-consistent mean-field approximation for Heisenberg Hamiltonian[20] has been carried out for an AFM-FM bilayer with a compensated interface. In accordance with the analytical model[18] the periodic modulation of the spin structure near the interface has been revealed. It is shown that the perpendicular orientation of the unit FM and AFM vectors at the interface is energetically favorable. This explains qualitatively the existence of the perpendicular exchange coupling in the bilayers $FeF_2$/Fe, $MnF_2$/Fe[12-14] and $Fe_3O_4$/CoO.[15-17] However, for AFM – FM bilayer with a uniform AFM film the transverse instability[19] has been observed within the AFM film at a critical value of external magnetic field. This instability breaks the initial exchange coupling at the interface, so that the resulting hysteresis loop is symmetrical. On the other hand, a shifted hysteresis loop with a finite exchange bias field has been obtained for a non-uniform AFM film consisting of various AFM domains with perpendicular directions of the easy anisotropy axes. This is because the exchange coupling at the interface never breaks completely if the AFM film consists of domains with sufficiently small lateral domain size at the interface.

## ACKNOWLEDGEMENT

This work was supported by means of ISTC grant #1991.